\documentstyle[12pt]{article}

\begin{document}

\newcommand{\dfrac}[2]{\displaystyle{\frac{#1}{#2}}}

{\it University of Shizuoka}

\hspace*{9.5cm} {\bf US-97-04}\\[-.3in]

\hspace*{9.5cm} {\bf June 1997}\\[.3in]

\vspace*{.4in}

\begin{center}

{\large\bf  Abnormal Structure of Fermion Mixings }\\[.2in]

{\large\bf  in a Seesaw Quark Mass Matrix Model  }\\[.3in]

{\bf Yoshio Koide}\footnote{
E-mail: koide@u-shizuoka-ken.ac.jp} \\

Department of Physics, University of Shizuoka \\ 
395 Yada, Shizuoka 422, Japan \\[.1in]

\vspace{.3in}

{\large\bf Abstract}\\[.1in]

\end{center}

\begin{quotation}
It is pointed out that in a seesaw quark mass matrix model which yields 
a singular enhancement of the top-quark mass, the right-handed 
fermion-mixing matrix $U_R^u$ for the up-quark sector has 
a peculiar structure in contrast to the left-handed one $U_L^u$.
As an example of the explicit structures of $U_L^u$ and $U_R^u$,
a case in which the heavy fermion mass matrix $M_F$ is given by 
a form [(unit matrix)+(rank-one matrix)] is investigated.
As a consequence, one finds observable signatures at projected 
high energy accelerators like the production of a fourth heavy 
quark family.
\end{quotation}

\vfill
PCAC numbers: 12.15.Ff, 13.15.Dk, 13.15.Jr.

\newpage

\centerline{\bf I. INTRODUCTION}

\vglue.05in

Why is the top-quark mass $m_t$ so singularly enhanced compared with the 
bottom-quark mass $m_b$ (while keeping $m_u\sim m_d$)?
Why does only the top-quark have a mass of the order of the 
electroweak scale $m_W$? 
Recently, it has been pointed out [1,2] that a seesaw quark mass matrix 
model [3] can give a natural answer to these questions.

In the conventional seesaw mass matrix model [3], 
we assume vector-like heavy fermions $F_i$ in addition to 
the conventional three-family quarks and leptons $f_i$ 
($f=u, d, \nu, e$;  $i=1, 2, 3$).
These fermions $f$ and $F$ belong to $f_L=(2,1)$, $f_R=(1,2)$, 
$F_L=(1,1)$ and $F_R=(1,1)$ of SU(2)$_L\times$SU(2)$_R$. 
The fermions $F$ acquire masses $M_F$ at a large energy scale 
$\mu\sim \lambda m_0$. 
The symmetries SU(2)$_L$ and SU(2)$_R$ are broken at the energy 
scales $\mu\sim m_0$ and $\mu\sim \kappa m_0$, respectively.
Then, the $6\times 6$ mass matrix $M$ for the fermions ($f, F$) 
is given by 
$$
M=\left( \begin{array}{cc}
0 & m_L \\
m_R & M_F \\ 
\end{array} \right) 
= m_0\left( \begin{array}{cc}
0 & Z_L \\
\kappa Z_R & \lambda Y \\ 
\end{array} \right) \ \ , 
\eqno(1.1)
$$
where matrices $Z_L$, $Z_R$ and $Y$ are dimensionless matrices 
with the order of one.
For $|\lambda|\gg |\kappa|\gg 1$ and det$M_F\neq 0$, the 
$3\times 3$ mass matrix for fermions $f$, $M_f$, is approximately 
given by the well-known ``seesaw" expression [4]
$$
M_f \simeq -m_L M_F^{-1} m_R \ , \eqno(1.2)
$$
so that the fermion masses $m^f_i$ ($i=1,2,3$) are suppressed by 
a factor $\kappa/\lambda$ to the electroweak scale $m_0$.
This was one of the motivations for considering a seesaw mechanism 
for quarks [3] before the discovery of the top quark [5]. 
However, the observation of the large top-quark mass  
has demanded that top-quark mass should be of the order of $m_0$ 
without the factor $\kappa/\lambda$.

Recently, it has been found [1,2] that the seesaw mass matrix with 
det$M_F=0$ can yield fermion masses $m_i^f$ and 
$m_i^F\equiv m_{i+3}^f$ $(i=1,2,3)$ with the following order:
$$
\begin{array}{l}
m_1, \ m_2 \sim (\kappa/\lambda) m_0 \ , \\
m_3 \sim m_0 \sim O(m_L) \ , \\
m_4 \sim \kappa m_0 \sim O(m_R) \ , \\
m_5, \ m_6 \sim \lambda m_0 \sim O(M_F) \ .
\end{array} \eqno(1.3)
$$
Note that the third fermion mass does not have the factor $\kappa/\lambda$.
Therefore, if the heavy fermion mass matrix $M_F$ takes det$M_F=0$ 
in the up-quark sector, we can understand 
why only the top-quark has a mass of the order of $m_L$ without the 
suppression factor $\kappa/\lambda$. 
This was first explicitly derived by Fusaoka and the author [1] 
on the basis of a special seesaw mass matrix model, ``democratic 
seesaw mass matrix model", where $M_F$ is given by the form 
[(unit matrix)+(a rank-one matrix)], and then generalized by Morozumi 
{\it et al.} [2].

In the present paper, we will point out that in such a model 
the right-handed fermion-mixing matrix $U_R$ has a peculiar structure 
in contrast to the left-handed one $U_L$, i.e., as if 
the third and fourth rows of $U_R$ are exchanged each other 
in contrast to $U_L$.
In Sec.~II, we will discuss general properties of the fermion 
mass spectrum and the mixing matrices $U_L$ and $U_R$ in the 
would-be seesaw mass matrix (1.1) with det$M_F=0$. 
In Sec.~III, in order to see the more explicit relations between 
the Cabibbo-Kobayashi-Maskawa (CKM) [6] matrices $V_L$ and $V_R$, 
we investigate a case with constraints $Z_R^T=Z_L$ and $Y^T=Y$ 
which are not so restrictive and which most models can satisfy. 
In Sec.~IV, as an explicit example of $U_L$ and $U_R$, 
we evaluate the mixing matrices for the case of the ``democratic 
seesaw mass matrix model"  which  can yield realistic quark masses 
and CKM matrix $V_L$.
The final section V will be devoted to the summary.

\vglue.2in
\centerline{\bf II. GENERAL STRUCTURES OF} 

\centerline{\bf THE FERMION MIXING MATRICES}

\vglue.05in

The mixing matrices $U_L$ and $U_R$ are obtained by diagonalizing 
the following Hermitian matrices $H_L$ and $H_R$, respectively:
$$
H_L \equiv M M^\dagger = \left(
\begin{array}{cc}
m_L m_L^\dagger & m_L M_F^\dagger \\
M_F m_L^\dagger & M_F M_F^\dagger + m_R m_R^\dagger 
\end{array} \right)
$$
$$
 = m_0^2 \left(
\begin{array}{cc}
Z_L Z_L^\dagger & \lambda Z_L Y^\dagger \\
\lambda Y Z_L^\dagger & \lambda^2 Y Y^\dagger + \kappa^2 Z_R Z_R^\dagger 
\end{array} \right) \ , \eqno(2.1) 
$$
$$
H_R \equiv M^\dagger M = \left(
\begin{array}{cc}
m_R^\dagger m_R & m_R^\dagger M_F \\
M_F^\dagger m_R &  M_F^\dagger M_F +  m_L^\dagger m_L
\end{array} \right)
$$
$$
 = m_0^2 \left(
\begin{array}{cc}
\kappa^2 Z_R^\dagger Z_R & \kappa \lambda Z_R^\dagger Y \\
\kappa\lambda Y^\dagger Z_R & \lambda^2 Y^\dagger Y + Z_L^\dagger Z_L
\end{array} \right) \ . \eqno(2.2) 
$$
For det$M_F\neq 0$, we can obtain the well-known seesaw expression (1.2)
since the $3\times 3$ Hermitian matrices $H_L^f$ and $H_R^f$ for 
fermions $f_i$ are approximately given by 
$$
H_L^f \simeq m_L m_L^\dagger - m_L M_F^\dagger (M_F M_F^\dagger 
+m_R m_R^\dagger)^{-1} M_F m_L^\dagger \simeq -m_L M_F^{-1} m_R
(m_L M_F^{-1} m_R)^\dagger \ , \eqno(2.3)
$$
$$
H_R^f \simeq  m_R^\dagger m_R - m_R^\dagger M_F (M_F^\dagger M_R
+m_L^\dagger m_L)^{-1} M_F^\dagger m_R\simeq  -(m_L M_F^{-1} m_R)^\dagger
m_L M_F^{-1} m_R \ . \eqno(2.4)
$$  
The $6\times 6$ mixing matrices $U_L$ and $U_R$ are 
approximately given by 
$$
U_L \simeq \left(\begin{array}{cc}
A_L & 0 \\ 
0 & B_L \\ 
\end{array} \right) 
\left(\begin{array}{cc}
{\bf 1} & -m_L M_F^{-1} \\ 
M_F^{\dagger -1} m_L^\dagger & {\bf 1} \\ 
\end{array} \right) 
= \left( \begin{array}{cc}
A_L & -A_L m_L M_F^{-1} \\ 
B_L M_F^{\dagger -1} m_L^\dagger & B_L \\ 
\end{array} \right) \ \ , 
\eqno(2.5)
$$
$$
U_R \simeq \left(\begin{array}{cc}
A_R & 0 \\ 
0 & B_R \\ 
\end{array} \right) 
\left(\begin{array}{cc}
{\bf 1} & -m_R^\dagger M_F^{\dagger -1} \\ 
M_F^{-1} m_R & {\bf 1} \\ 
\end{array} \right) 
= \left(\begin{array}{cc}
A_R & -A_R m_R^\dagger M_F^{\dagger -1} \\ 
B_R M_F^{-1} m_R & B_R \\ 
\end{array} \right) \ \ , 
\eqno(2.6)
$$
where the $3\times 3$ unitary matrices $A$ and $B$ are defined by
$$
-A_L m_L M_F^{-1} m_R A_R^\dagger = D_f \  , \ \ \ 
B_L M_F B_R^\dagger = D_F \ \ , 
\eqno(2.7)
$$
$D_f = {\rm diag}(m_1^f, m_2^f, m_3^f)$ and 
$D_F = {\rm diag}(m_4^f, m_5^f, m_6^f) \equiv 
{\rm diag}(m_1^F, m_2^F, m_3^F)$. 
The mixing matrix $U_R$ has a structure similar to $U_L$ 
except for the point that the off-diagonal elements $(U_L)_{ik}$ 
and $(U_L)_{ki}$ ($i=1,2,3$; $k=4,5,6$) have a suppression factor 
$1/\lambda$, while $(U_R)_{ik}$ and $(U_R)_{ki}$ have a suppression 
factor $\kappa/\lambda$.

On the other hand, for the case of det$M_F=0$, the seesaw expression
(1.2) is not valid any longer.
For the case of det$M_F=0$, without losing generality, we can 
choose a heavy fermion basis where the mass matrix $M_F$ is 
given by a diagonal form 
$$
M_F = \lambda m_0 \left(
\begin{array}{ccc}
0 & 0 & 0 \\
0 & \ast & 0 \\
0 & 0 & \ast 
\end{array} \right) \ , \eqno(2.8)
$$
where $\ast$ denote  elements with the order of one.
Then the Hermitian matrices (2.1) and (2.2) take the following 
textures:
$$ 
H_L = m_0^2 \left(
\begin{array}{cccccc}
\ast & \ast & \ast & 0  & \sim\lambda &  \sim\lambda \\
\ast & \ast & \ast & 0  & \sim\lambda &  \sim\lambda \\
\ast & \ast & \ast & 0  & \sim\lambda &  \sim\lambda \\
0    & 0    & 0    & \sim\kappa^2 & \sim\kappa^2 & \sim\kappa^2 \\
\sim\lambda & \sim\lambda & \sim\lambda 
& \sim\kappa^2 & \sim\lambda^2 & \sim\kappa^2 \\
\sim\lambda & \sim\lambda & \sim\lambda 
& \sim\kappa^2 & \sim\kappa^2 & \sim\lambda^2  \\
\end{array} \right) \ , \eqno(2.9)
$$
$$ 
H_R = m_0^2 \left(
\begin{array}{cccccc}
\sim\kappa^2 & \sim\kappa^2 & \sim\kappa^2 
& 0  & \sim\kappa\lambda &  \sim\kappa\lambda \\
\sim\kappa^2 & \sim\kappa^2 & \sim\kappa^2 
& 0  & \sim\kappa\lambda &  \sim\kappa\lambda \\
\sim\kappa^2 & \sim\kappa^2 & \sim\kappa^2 
& 0  & \sim\kappa\lambda &  \sim\kappa\lambda \\
0    & 0    & 0    & \ast & \ast & \ast \\
\sim\kappa\lambda & \sim\kappa\lambda & \sim\kappa\lambda 
& \ast & \sim\lambda^2 & \ast \\
\sim\kappa\lambda & \sim\kappa\lambda & \sim\kappa\lambda 
& \ast & \ast & \sim\lambda^2  \\
\end{array} \right) \ . \eqno(2.10)
$$
Note that $(H_L)_{33}\ll (H_L)_{44}$, while 
$(H_R)_{33}\gg (H_R)_{44}$. 
This causes the exchange between the third and fourth rows in $U_R$
in contrast to $U_L$.
As a result, the mixing matrix $U_R$ has matrix elements of the order 
$$
U_R =  \left(
\begin{array}{cccccc}
\ast & \ast & \ast & \sim\frac{\kappa}{\lambda}  & 
\sim\frac{\kappa}{\lambda} & \sim\frac{\kappa}{\lambda} \\
\ast & \ast & \ast & \sim\frac{\kappa}{\lambda}  & 
\sim\frac{\kappa}{\lambda} & \sim\frac{\kappa}{\lambda} \\
\sim\frac{\kappa}{\lambda} & \sim\frac{\kappa}{\lambda} 
& \sim\frac{\kappa}{\lambda} & \ast & \ast & \ast \\
\ast & \ast & \ast & \sim\frac{\kappa}{\lambda}  & 
\sim\frac{\kappa}{\lambda} & \sim\frac{\kappa}{\lambda} \\
\sim\frac{\kappa}{\lambda} & \sim\frac{\kappa}{\lambda} 
& \sim\frac{\kappa}{\lambda} & \ast & \ast & \ast \\
\sim\frac{\kappa}{\lambda} & \sim\frac{\kappa}{\lambda} 
& \sim\frac{\kappa}{\lambda} & \ast & \ast & \ast \\
\end{array} \right) \  , \eqno(2.11)
$$
in contrast to 
$$
U_L =  \left(
\begin{array}{cccccc}
\ast & \ast & \ast & \sim\frac{1}{\lambda}  & 
\sim\frac{1}{\lambda} & \sim\frac{1}{\lambda} \\
\ast & \ast & \ast & \sim\frac{1}{\lambda}  & 
\sim\frac{1}{\lambda} & \sim\frac{1}{\lambda} \\
\ast & \ast & \ast & \sim\frac{1}{\lambda}  & 
\sim\frac{1}{\lambda} & \sim\frac{1}{\lambda} \\
\sim\frac{1}{\lambda} & \sim\frac{1}{\lambda} 
& \sim\frac{1}{\lambda} & \ast & \ast & \ast \\
\sim\frac{1}{\lambda} & \sim\frac{1}{\lambda} 
& \sim\frac{1}{\lambda} & \ast & \ast & \ast \\
\sim\frac{1}{\lambda} & \sim\frac{1}{\lambda} 
& \sim\frac{1}{\lambda} & \ast & \ast & \ast \\
\end{array} \right) \  . \eqno(2.12)
$$
The structures (2.11) and (2.12) mean that the dominant 
components of the fermions in the up-quark sector are 
given by
$$
\begin{array}{ll}
u_L \simeq (2,1) \ , \ \ \ & u_R \simeq (1,2) \ , \\
c_L \simeq (2,1) \ , \ \ \ & c_R \simeq (1,2) \ , \\
t_L \simeq (2,1) \ , \ \ \ & t_R \simeq (1,1) \ , \\
t'_L \simeq (1,1) \ , \ \ \ & t'_R \simeq (1,2) \ , \\
u_{5L} \simeq (1,1) \ , \ \ \ & u_{5R}\simeq (1,1) \ , \\
u_{6L} \simeq (1,1) \ , \ \ \ & u_{6R}\simeq (1,1) \ , \\
\end{array} \eqno(2.13)
$$
of SU(2)$_L\times$U(2)$_R$, where we have denoted the fermion 
$u_4$ as $t'$. 
We should notice that $t$ and $t'$ have exceptional structures 
differently from other fermions $f$ and $F$.

We consider that in the down-quark sector the seesaw expression 
(1.2) is well satisfied, so that the mixing matrices $U_L$ and $U_R$ 
are given by normal structure as (2.12). 
Then, the CKM matrix $V_L$ is  given by 
$$
V_L =  \left(
\begin{array}{cccccc}
\ast & \ast & \ast & \sim\frac{1}{\lambda}  & 
\sim\frac{1}{\lambda} & \sim\frac{1}{\lambda} \\
\ast & \ast & \ast & \sim\frac{1}{\lambda}  & 
\sim\frac{1}{\lambda} & \sim\frac{1}{\lambda} \\
\ast & \ast & \ast & \sim\frac{1}{\lambda}  & 
\sim\frac{1}{\lambda} & \sim\frac{1}{\lambda} \\
\sim\frac{1}{\lambda} & \sim\frac{1}{\lambda} 
& \sim\frac{1}{\lambda} & \sim\left(\frac{1}{\lambda}\right)^2 
& \sim\left(\frac{1}{\lambda}\right)^2 & 
\sim\left(\frac{1}{\lambda}\right)^2 \\
\sim\frac{1}{\lambda} & \sim\frac{1}{\lambda} 
& \sim\frac{1}{\lambda} & \sim\left(\frac{1}{\lambda}\right)^2 
& \sim\left(\frac{1}{\lambda}\right)^2 
& \sim\left(\frac{1}{\lambda}\right)^2 \\
\sim\frac{1}{\lambda} & \sim\frac{1}{\lambda} 
& \sim\frac{1}{\lambda} & \sim\left(\frac{1}{\lambda}\right)^2 
& \sim\left(\frac{1}{\lambda}\right)^2 
& \sim\left(\frac{1}{\lambda}\right)^2 \\
\end{array} \right) \  , \eqno(2.14)
$$
while the CKM matrix $V_R$ for the right-handed weak currents is 
given by 
$$
V_R =  \left(
\begin{array}{cccccc}
\ast & \ast & \ast & \sim\frac{\kappa}{\lambda}  & 
\sim\frac{\kappa}{\lambda} & \sim\frac{\kappa}{\lambda} \\
\ast & \ast & \ast & \sim\frac{\kappa}{\lambda}  & 
\sim\frac{\kappa}{\lambda} & \sim\frac{\kappa}{\lambda} \\
\sim\frac{\kappa}{\lambda} & \sim\frac{\kappa}{\lambda} 
& \sim\frac{\kappa}{\lambda} 
& \sim\left(\frac{\kappa}{\lambda}\right)^2 
& \sim\left(\frac{\kappa}{\lambda}\right)^2 
& \sim\left(\frac{\kappa}{\lambda}\right)^2 \\
\ast & \ast & \ast & \sim\frac{\kappa}{\lambda}  & 
\sim\frac{\kappa}{\lambda} & \sim\frac{\kappa}{\lambda} \\
\sim\frac{\kappa}{\lambda} & \sim\frac{\kappa}{\lambda} 
& \sim\frac{\kappa}{\lambda} 
& \sim\left(\frac{\kappa}{\lambda}\right)^2 
& \sim\left(\frac{\kappa}{\lambda}\right)^2 
& \sim\left(\frac{\kappa}{\lambda}\right)^2 \\
\sim\frac{\kappa}{\lambda} & \sim\frac{\kappa}{\lambda} 
& \sim\frac{\kappa}{\lambda} 
& \sim\left(\frac{\kappa}{\lambda}\right)^2 
& \sim\left(\frac{\kappa}{\lambda}\right)^2 
& \sim\left(\frac{\kappa}{\lambda}\right)^2 \\
\end{array} \right) \  , \eqno(2.15)
$$
where the factors $(1/\lambda)^2$ and $(\kappa/\lambda)^2$ come from 
the reason that the heavy fermions $F_i$ are $(1,1)$ of 
SU(2)$_L\times$SU(2)$_R$ in the present model. 

\vspace{.2in}
\centerline
{\bf III. CKM MATRICES $V_L$ AND $V_R$}

\vglue.05in

In order to see these relations (2.14) and (2.15) explicitly, we 
consider a model with additional constraint 
$$
Z_R^T=Z_L \ , \ \ \ Y^T=Y \ . \eqno(3.1)
$$
The constraint (3.1) are not so restrictive, and most seesaw mass 
matrix model will satisfy this constraint.

For the down-quark sector in which the seesaw expression (1.2) 
is valid, from (2.1) -- (2.6), we obtain the relations 
$$
\begin{array}{ll}
U^R_{dd}\simeq (U^L_{dd})^* , \ \ & 
U^R_{dD}\simeq \kappa (U^L_{dD})^* \ , \\
U^R_{Dd}\simeq \kappa (U^L_{Dd})^* \ , \ & 
U^R_{DD}\simeq (U^L_{DD})^* , \ \\
\end{array} \eqno(3.2)
$$
where the $3\times 3$ matrices $U_{ab}$  ($a,b=f,F$) 
are defined by
$$
U^f = \left(
\begin{array}{cc}
U_{ff} & U_{fF} \\
U_{Ff} & U_{FF} \\
\end{array} \right) \ . \eqno(3.3)
$$

For the up-quark sector with det$M_F=0$, 
the seesaw expression (1.2) [therefore, (2.5) and (2.6)]
is not valid any longer.
However, when we define $\widetilde{U}_R=P_{34}U_R$, 
where
$$
P_{34}=\left(
\begin{array}{cccccc}
1 & 0 & 0 & 0 & 0 & 0 \\
0 & 1 & 0 & 0 & 0 & 0 \\
0 & 0 & 0 & 1 & 0 & 0 \\
0 & 0 & 1 & 0 & 0 & 0 \\
0 & 0 & 0 & 0 & 1 & 0 \\
0 & 0 & 0 & 0 & 0 & 1 \\
\end{array} \right) \ , \eqno(3.4)
$$
we can see that the mixing matrix $\widetilde{U}_R$ has a 
structure similar to $U_L$, because $\widetilde{U}_R H_R 
\widetilde{U}_R^\dagger =(DP_{34})^\dagger DP_{34} = 
{\rm diag}(m_1^2, m_2^2, m_4^2, m_3^2, m_5^2, m_6^2)$
has the structure similar to $U_L H_L U_L^\dagger = D^2=
{\rm diag}(m_1^2, m_2^2, m_3^2, m_4^2, m_5^2, m_6^2)$ 
apart from the exchange of the coefficients, 
$1\leftrightarrow\kappa$ [see (2.9) and (2.10)]. 
Therefore, we obtain the relations 
$$
\begin{array}{ll}
\widetilde{U}^R_{uu}\simeq (U^L_{uu})^* , \ \ & 
\widetilde{U}^R_{uU}\simeq \kappa (U^L_{uU})^* \ , \\
\widetilde{U}^R_{Uu}\simeq \kappa (U^L_{Uu})^* \ , \ & 
\widetilde{U}^R_{UU}\simeq (U^L_{UU})^* , \ \\
\end{array} \eqno(3.5)
$$
similarly to (3.2).

Since the CKM mixing matrix $V_L$ for the left-handed weak currents is 
given by 
$$
V_L=\left(
\begin{array}{cc}
U_{uu}^L & U_{uU}^L \\
U_{Uu}^L & U_{UU}^L \\
\end{array} \right) \left(
\begin{array}{cc}
{\bf 1} & 0 \\
0 & 0 \\
\end{array} \right)  \left(
\begin{array}{cc}
U_{dd}^{L\dagger} & U_{Dd}^{L\dagger} \\
U_{dD}^{L\dagger} & U_{DD}^{L\dagger} \\
\end{array} \right)
 = \left(
\begin{array}{cc}
U_{uu}^LU_{dd}^{L\dagger} & U_{uu}^LU_{Dd}^{L\dagger} \\
U_{Uu}^LU_{dd}^{L\dagger} & U_{Uu}^LU_{Dd}^{L\dagger} \\
\end{array} \right) \ , \eqno(3.6)
$$
the CKM mixing matrix $V_R$ for the right-handed weak currents is 
given by 
$$
V_R= P_{34} \left(
\begin{array}{cc}
\widetilde{U}_{uu}^R & \widetilde{U}_{uU}^R \\
\widetilde{U}_{Uu}^R & \widetilde{U}_{UU}^R \\
\end{array} \right) \left(
\begin{array}{cc}
{\bf 1} & 0 \\
0 & 0 \\
\end{array} \right)  \left(
\begin{array}{cc}
U_{dd}^{R\dagger} & U_{Dd}^{R\dagger} \\
U_{dD}^{R\dagger} & U_{DD}^{R\dagger} \\
\end{array} \right) =  P_{34} \left(
\begin{array}{cc}
\widetilde{U}_{uu}^RU_{dd}^{R\dagger} & 
\widetilde{U}_{uu}^RU_{Dd}^{R\dagger} \\
\widetilde{U}_{Uu}^RU_{dd}^{R\dagger} & 
\widetilde{U}_{Uu}^RU_{Dd}^{R\dagger} \\
\end{array} \right) 
$$
$$
=  P_{34} \left(
\begin{array}{cc}
(V_{ud}^L)^* & \kappa (V_{uD}^L)^* \\
\kappa (V_{Ud}^L)^* & \kappa^2 (V_{UD}^L)^* \\
\end{array} \right) \ . \eqno(3.7)
$$
Therefore, we find
$$
\begin{array}{ll}
V_{ij}^R = (V_{ij}^L)^* & i=u,c; \ j=d,s,b \ , \\
V_{tj}^R = (V_{t'j}^L)^* &  j=d,s,b \ , \\
V_{t'j}^R = (V_{tj}^L)^* &  j=d,s,b \ . \\
\end{array} \eqno(3.8)
$$
As seen from (3.8), the right-handed weak-interaction structure 
of $t'$ is the same as the left-handed weak interaction 
structure of $t$.

In general, in the left-right symmetric model [7], 
the $W_R$-exchange diagrams  can sizably contribute to the 
$K^0$-$\overline{K}^0$ mixing [8]. 
However, in the present model, although the right-handed weak 
currents for $u$ and $c$ can contribute to the 
$K^0$-$\overline{K}^0$ mixing as pointed out 
in Ref.~[7], those for $t$ and $t'$ are negligibly 
small, because $t_L$ ($t'_R$) is doublet of SU(2)$_{L(R)}$, 
while $t_R$ ($t'_L$) is almost singlet of SU(2)$_{R(L)}$. 
Also the contributions for $u_5$ and $u_6$ are negligibly 
small because of the suppression factors $(1/\lambda)^2$ and
$(\kappa/\lambda)^2$.
For example, the $K^0$-$\overline{K}^0$ mixing amplitude 
via $(t,t', W_L, W_R)$ is suppressed by a factor $(1/\lambda)^2$ 
compared with that via $(t,t, W_L, W_L)$. 
The next leading term to the diagram $(t,t, W_L, W_L)$ is 
a diagram $(t',t', W_R, W_R)$ which is suppressed by a factor 
$(1/\kappa)^2$ compared with the diagram $(t,t, W_L, W_L)$.
For $\kappa \geq 10$, the contributions are negligibly small.

Since the fourth up-quark $t'$ has a comparatively light mass
(of the order of $m_{W_R}$), we can expect the observation of 
$t'$-production via the reaction $d+u\rightarrow t'+d$ 
with the $W_R$ exchange, for example, at LHC. 
Since we consider $m(W_R)\simeq \kappa m(W_L)$, we obtain 
$$
\sigma(p+p\rightarrow t'+X) \simeq \frac{1}{\kappa^4}
\sigma(p+p\rightarrow t+X) \ . \eqno(3.9)
$$
The decay width of $t'$, $\Gamma_{t'}$, is given by 
$$
\frac{\Gamma_{t'}}{\Gamma_t} \simeq 
\frac{m_{t'}^5/m_{W_R}^4}{m_{t}^5/m_{W_L}^4}
\simeq \kappa \ , \eqno(3.10)
$$
from (3.8).

\vglue.2in

\centerline{\bf IV.  NUMERICAL EXAMPLE FOR A SPECIFIC MODEL}

\vglue.05in

Thus, as far as the mass matrix (1.1) satisfies the form (3.1),
the CKM matrix $V_R$ can be related to $V_L$ 
irrelevantly to the explicit structures of $Z$ and $Y$.
However, in order to see the effects of the flavor-changing
neutral currents (FCNC), we need the explicit forms of 
$U_{L(R)}^u$ and $U_{L(R)}^d$ separately, because 
the left- (right-) handed FCNC among the conventional fermions 
$f_i$ ($i=1,2,3$) are proportional [9] to the matrices 
(see Appendix)
$$
C_{L(R)}^f = U_{fF}^{L(R)} ( U_{fF}^{L(R)})^\dagger 
\ . \eqno(4.1)
$$
So, it is interesting to see the explicit structures of 
$U_L$ and $U_R$ for a realistic model which can give 
reasonable quark masses and the CKM mixing matrix parameters.

As an example, we choose the democratic seesaw mass matrix 
model [1],  where $M_F$ is given by the form [(unit matrix)
+(a rank-one matrix)]:
$$
M_F = \lambda m_0 Y_f = \lambda m_0 ({\bf 1} + 3 b_f X) \  , 
\eqno(4.2)
$$
where {\bf 1} and $X$ are the $3\times 3$ unit matrix and 
a rank-one matrix with the condition $X^2=X$, respectively, 
and $b_f$ is an $f$-dependent complex parameter. 
The name ``democratic" [10] comes from the following assumption:
the matrices $Z_L$ and $Z_R$ are given by a diagonal form 
in the heavy fermion basis on which the matrix $X$ is democratic, 
i.e.,
$$
X = \frac{1}{3}\left(
\begin{array}{ccc}
1 & 1 & 1 \\
1 & 1 & 1 \\
1 & 1 & 1 \\
\end{array} \right) \  ,\eqno(4.3)
$$ 
For simplicity, we assume that the matrices 
$Z_L$ and $Z_R$ have a common structure except for 
their phases:
$$
Z_L=P(\delta_L) Z \ , \ \ \ Z_R=P(\delta_R) Z \ , \eqno(4.4)
$$
$$
P(\delta)={\rm diag}(e^{i\delta_1}, e^{i\delta_2}, e^{i\delta_3}) \ , 
\eqno(4.5)
$$
where the matrix $Z$ is a real-parameter matrix 
$$
Z =  \left(
\begin{array}{ccc}
z_1 & 0 & 0 \\
0 & z_2 & 0 \\
0 & 0 & z_3 \\ 
\end{array} \right) \ \ , 
\eqno(4.6)
$$
with $z_1^2 + z_2^2 + z_3^2=1$, and it is universal for 
all fermion sectors (up- and down-, quark and lepton sectors).
In order to obtain input values for the parameters $z_i$, 
we assume that the parameter $b_f$ takes the value $b_e=0$
in the charged lepton sector, so that the parameters $z_i$ 
are given by 
$$
\frac{z_1}{\sqrt{m_e}} = \frac{z_2}{\sqrt{m_\mu}} = \frac{z_3}{\sqrt{m_\tau}} 
= \frac{1}{\sqrt{m_e + m_\mu + m_\tau}} 
\eqno(4.7)
$$
from $M_e = m_0(\kappa/\lambda)P(\delta_L^e -\delta_R^e)\cdot
 Z\cdot{\bf 1}\cdot Z$.
The  ansatz of the democratic $M_F$ was motivated by 
the successful relation [11]
$$
\frac{m_u}{m_c}\simeq \frac{3}{4}\frac{m_e}{m_\mu}  \ , \eqno(4.8)
$$
(independently of $\kappa/\lambda$ under $\lambda\gg \kappa$) for 
$b_u=-1/3$ and $b_e=0$.
In Ref.~[1], the value of $\kappa/\lambda$ has been fixed as 
$\kappa/\lambda=0.02$ by the relations for $b_u\simeq -1/3$ 
$$
m_c \simeq 2 \frac{m_\mu}{m_\tau} \frac{\kappa}{\lambda} m_0 \ , 
\ \ \ m_t \simeq \frac{1}{\sqrt{3}} m_0 \ , \eqno(4.9)
$$
(note that the expression of $m_t$ does not contain the suppression 
factor $\kappa/\lambda$).
The value of $\beta_d\equiv \arg(-b_d)$ has been chosen as 
$\beta_d = \pi/10$ from the relations for $b_d \simeq -1$
$$
m_s \simeq 2\left|\sin\frac{\beta_d}{2}\right| 
\frac{m_\mu}{m_\tau}\frac{\kappa}{\lambda} m_0\ , \ \ \ 
m_b \simeq \frac{1}{2} \frac{\kappa}{\lambda} m_0 
\ . \eqno(4.10)
$$
Then, we can find in Ref.~[1] that these parameter values 
can successfully provide all of the quark mass ratios and 
CKM matrix parameters.
It is worth while noting that the model can yield 
$m_t\gg m_b$ with keeping $m_u\sim m_d$ by adjusting 
only one complex parameter $b_f$ and without choosing 
hierarchically different values between $b_u$ and $b_d$.

In order to evaluate the CKM matrices $V_L$ and $V_R$, 
it is convenient to define the matrix $\overline{M}$ which 
are given by 
$$
\overline{M}=P^\dagger (\delta_L) M P^\dagger(\delta_R) 
= m_0 \left( 
\begin{array}{cc}
0 & Z \\
\kappa Z & \lambda Y \\
\end{array} \right) \ , 
\eqno(4.11)
$$
where the $6\times 6$ phase matrix $P(\delta)$ is defined by
$$
P(\delta) = {\rm diag}(e^{i\delta_1}, e^{i\delta_2}, e^{i\delta_3},
1,1,1) \ , \eqno(4.12)
$$
(we have used the same notation with the $3\times 3$ phase matrix 
(4.5)).
The unitary matrices $U_L$ and $U_R$ are related to 
$$
U_L= \overline{U}_L P^\dagger(\delta_L) \ , \ \ \ 
U_R= \overline{U}_R P(\delta_R) \ ,
\eqno(4.13)
$$
where
$$
U_L M U_R^\dagger = \overline{U}_L \overline{M}\, 
\overline{U}_R^\dagger = D \ . 
\eqno(4.14)
$$
Then, the CKM matrices $V_L$ and $V_R$ are given by
$$
\begin{array}{l}
V_L = U_L^u P_0 U_L^{d\dagger} = \overline{U}_L^u 
P(\delta_L^d -\delta_L^u) P_0 \overline{U}_L^{d\dagger}\ , \\
V_R = U_R^u P_0 U_R^{d\dagger} = \overline{U}_R^u 
P(\delta_R^u -\delta_R^d) P_0 \overline{U}_R^{d\dagger}\ , \\
\end{array} 
\eqno(4.15)
$$
where
$$
P_0 ={\rm diag}(1,1,1,0,0,0) \ . \eqno(4.16)
$$
Since the constraint (3.1) means 
$\delta^f_{Ri}=\delta^f_{Li}\equiv \delta^f_i$, 
so that $V_L$ and $V_R$ are given by
$$
V_L =  \overline{U}_L^u P_0(\delta) \overline{U}_L^{d\dagger} 
\ , \ \ \ 
V_R =  \overline{U}_R^u P_0^\dagger(\delta) \overline{U}_R^{d\dagger}
\ , \eqno(4.17)
$$
where 
$$
P_0(\delta)= {\rm diag}(e^{i\delta_1}, e^{i\delta_2}, 
e^{i\delta_3},0,0,0) 
\ , \eqno(4.18)
$$
with $\delta_i=-(\delta_i^u -\delta_i^d)$.

The observed CKM matrix parameters are roughly  described 
by $(\delta_1, \delta_2, \delta_3)=(0,0,\pi)$ [1]
and more precisely by 
$(\delta_1, \delta_2, \delta_3)=(0,0,\pi-\pi/30)$ [12]. 
However, in Refs.~[1] and [12], only the $3\times 3$ part of $V_L$ 
has been investigated. 
Here, we show the numerical results of the $6\times 6$ mixing 
matrices $U_L$ and $U_R$ for the case of $\kappa/\lambda=0.02$, 
$\beta_d=\pi/10$, and $\delta_3=\pi -\pi/30$ 
(we take $\kappa=10$ temporarily according to Ref.[1], 
but the results are 
almost insensitive to the value of $\kappa$):
$$
U_L^u = \left(
\begin{array}{lll|lll} 
+0.9994 & -0.0349 & -0.0084 & -0.0247 \frac{1}{\lambda} & 
+6\times 10^{-5}\frac{1}{\lambda} & +4\times 10^{-6}\frac{1}{\lambda} \\
+0.0319 & +0.9709 & -0.2373 & -0.2051\frac{1}{\lambda} 
& -0.4345\frac{1}{\lambda} & +0.0259\frac{1}{\lambda} \\
+0.0165 & +0.2369 & +0.9714 & +0.8989\frac{1}{\lambda} & 
+0.8431\frac{1}{\lambda} & -0.0444\frac{1}{\lambda} \\ 
\hline
+0.0093\frac{1}{\lambda} & +0.1114\frac{1}{\lambda} 
& -1.0364\frac{1}{\lambda} & +0.5774 & +0.5774 & +0.5772 \\
-0.0118\frac{1}{\lambda} & +0.1649\frac{1}{\lambda} 
& +0.0209\frac{1}{\lambda} & -0.7176 & +0.6961 & +0.0215 \\
-0.0064\frac{1}{\lambda} & -0.1011\frac{1}{\lambda} 
& +0.7927\frac{1}{\lambda} & -0.3894 & -0.4267 & +0.8163 
\end{array} \right) \ ,
\eqno(4.19)
$$
$$
U_R^u = \left(
\begin{array}{lll|lll} 
+0.9994 & -0.0349 & -0.0084 & -0.0247\frac{\kappa}{\lambda} & 
+6\times 10^{-5}\frac{\kappa}{\lambda} 
& +4\times 10^{-6}\frac{\kappa}{\lambda} \\
+0.0319 & +0.9709 & -0.2373 & -0.2051\frac{\kappa}{\lambda} 
& -0.4346\frac{\kappa}{\lambda} & +0.0259\frac{\kappa}{\lambda} \\
+0.0256\frac{\kappa}{\lambda} & +0.3459\frac{\kappa}{\lambda} 
& -0.0747\frac{\kappa}{\lambda} & +0.5773 & +0.5773 & +0.5774 \\
\hline
+0.0165 & +0.2369 & +0.9713 & +0.3274\frac{\kappa}{\lambda} 
& +0.2716\frac{\kappa}{\lambda} & -0.6159\frac{\kappa}{\lambda} \\ 
-0.0118\frac{\kappa}{\lambda} & +0.1649\frac{\kappa}{\lambda} 
& +0.0209\frac{\kappa}{\lambda} & -0.7176 & +0.6961 & +0.0215 \\
-0.0064\frac{\kappa}{\lambda} & -0.1010\frac{\kappa}{\lambda} 
& +0.7929\frac{\kappa}{\lambda} & -0.3894 & -0.4267 & +0.8161 
\end{array} \right) \ .
\eqno(4.20)
$$
$$
|U_L^d| = \left(
\begin{array}{lll|lll} 
0.9772 &  0.2061  &  0.0506 &  0.0490\frac{1}{\lambda} 
& 0.0007\frac{1}{\lambda} & 4\times 10^{-5}\frac{1}{\lambda} \\
 0.2118 & 0.9540  & 0.2124 & 0.2063\frac{1}{\lambda} 
&  0.0646\frac{1}{\lambda} &  0.0035\frac{1}{\lambda} \\
0.0137 & 0.2179  & 0.9759 & 0.4335\frac{1}{\lambda} 
& 0.4809\frac{1}{\lambda} &  0.5251\frac{1}{\lambda} \\ 
\hline
0.0118\frac{1}{\lambda} & 0.1649\frac{1}{\lambda} & 0.0209\frac{1}{\lambda} 
& 0.7176 & 0.6961 & 0.0215 \\
0.0064\frac{1}{\lambda} & 0.1010\frac{1}{\lambda} & 0.7927\frac{1}{\lambda} 
& 0.3895 & 0.4268 & 0.8162 \\
0.0046\frac{1}{\lambda} & 0.0660\frac{1}{\lambda} & 0.2706\frac{1}{\lambda} 
& 0.5773 & 0.5773 & 0.5774 
\end{array} \right) \ ,
\eqno(4.21)
$$
where for $U_L^d$, for simplicity, we have shown only the magnitudes. 
Since the mixing matrix elements of $U_R^d$ are given by the relation 
(3.5) with good approximation, here we have dropped the numerical 
result of $U_R^d$. 

{}From (4.17), the $6\times 6$ CKM matrix $V_L$ is given by 
$$
|V_L| = \left(
\begin{array}{lll|lll} 
0.9756 &  0.2196  &  0.0028 &  0.0174\frac{1}{\lambda} 
& 0.0038\frac{1}{\lambda} & 0.0046\frac{1}{\lambda} \\
0.2193 & 0.9749  & 0.0388 & 0.1615\frac{1}{\lambda} 
&  0.0910\frac{1}{\lambda} &  0.1283\frac{1}{\lambda} \\
0.0105 & 0.0374  & 0.9992 & 0.0188\frac{1}{\lambda} 
& 0.7940\frac{1}{\lambda} &  0.2473\frac{1}{\lambda} \\ 
\hline
0.0780\frac{1}{\lambda} & 0.3253\frac{1}{\lambda} & 0.9873\frac{1}{\lambda} 
& 00399\left(\frac{1}{\lambda}\right)^2 & 
0.8104\left(\frac{1}{\lambda}\right)^2 & 
0.2879\left(\frac{1}{\lambda}\right)^2 \\
0.0332\frac{1}{\lambda} & 0.1534\frac{1}{\lambda} & 0.0560\frac{1}{\lambda} 
& 0.0269\left(\frac{1}{\lambda}\right)^2 & 
0.0331\left(\frac{1}{\lambda}\right)^2 & 
0.0052\left(\frac{1}{\lambda}\right)^2 \\
0.0626\frac{1}{\lambda} & 0.2638\frac{1}{\lambda} & 0.7517\frac{1}{\lambda} 
& 0.0331\left(\frac{1}{\lambda}\right)^2 & 
0.6182\left(\frac{1}{\lambda}\right)^2 & 
0.2212\left(\frac{1}{\lambda}\right)^2 
\end{array} \right) \ .
\eqno(4.22)
$$
We have again dropped the results of $V_R$ since the numerical 
results satisfies (3.7) very well.

The numerical results of $C_L$ and $C_R$, to which the contributions 
of FCNC are proportional, are given as follows:
$$
C_L^u = \left(
\begin{array}{lll} 
2.43314\times 10^{-9} &  -2.013\times 10^{-8}  & 
-8.845\times 10^{-8} \\
-2.013\times 10^{-8} & 9.263\times 10^{-7}  & 
2.208\times 10^{-6} \\
-8.848\times 10^{-8} & 2.208\times 10^{-6} & 
6.084\times 10^{-6} 
\\
\end{array} \right) \ ,
\eqno(4.23)
$$
$$
C_R^u = \left(
\begin{array}{lll} 
2.43314\times 10^{-7} &  2.013\times 10^{-6}  & 
0.0002840 \\
2.013\times 10^{-6} & 9.264\times 10^{-5}  & 0.007087 \\
0.0002840  & 0.007087 & 1.000 
\end{array} \right) \ ,
\eqno(4.24)
$$
$$
|C_L^d| = \frac{1}{\kappa^2} |C_R^d| = \left(
\begin{array}{lll} 
9.615\times 10^{-9} &  4.026\times 10^{-8}  & 
8.525\times 10^{-8} \\
4.026\times 10^{-8} & 1.870\times 10^{-7}  & 
3.514\times 10^{-7} \\
8.525\times 10^{-8} & 3.514\times 10^{-7} & 
2.780\times 10^{-6} \\
\end{array} \right) \ .
\eqno(4.25)
$$
Thus, the matrix elements of $C_L$ and $C_R$ are suppressed by  factors 
$(1/\lambda)^2$ and $(\kappa/\lambda)^2$, respectively, except for 
$(C_R^u)_{3i}=(C_R^u)_{i3}$ ($i=1,2,3$).
We see that the  FCNC in the present model are harmless 
to the $K^0$-$\overline{K}^0$ and $D^0$-$\overline{D}^0$ mixings.
However, the elements related to the top-quark have sizable values 
of $C_R^u$: 
$$
(C_R^u)_{tc}=0.000709 \ , \ \ \ \    (C_R^u)_{tu}=0.000284 \ .
\eqno(4.26)
$$
The observability of the single top-quark productions via FCNC with 
(4.26) will be discussed elsewhere [9].

\vglue.2in

\centerline{\bf V. SUMMARY}
\vglue.05in

In conclusion, we have pointed out that in a seesaw quark mass matrix 
model which yields a singular enhancement of the top-quark mass, 
the $6\times 6$ mixing matrix $U_R^u$ for the right-handed up-quark 
sector has a peculiar structure, i.e., as if the third and fourth rows 
of $U_R^u$ are exchanged in contrast to the left-handed mixing matrix 
$U_L^u$. 
This means that top quark $t$  and the fourth up-quark $t'$ have 
approximately components $t_L=(2,1)$, $t_R=(1,1)$, $t'_L=(1,1)$ and 
$t'_R=(1,2)$ of SU(2)$_L\times$SU(2)$_R$, although other fermions have 
$f_L=(2,1)$, $f_R=(1,2)$, $F_L=(1,1)$ and $F_R=(1,1)$.

For a model with the constraint (3.1) which is a likely case, 
the CKM mixing matrices $V_L$ and $V_R$ satisfy the relation (3.8).
Observation of $t'$ with mass $m_{t'}\simeq \kappa m_t$ (of the 
order of $m_{W_R}$) is expected at a future collider with a few 
TeV energy.

As an explicit example of $U_L$ and $U_R$, we have investigated 
a model where $M_F$ is given by a form (4.2). 
The numerical results, of course,  satisfy the general relations 
(3.5) and (3.7).
The matrix elements of $C=U_{fF} (U_{fF})^\dagger$ to which
FCNC are proportional have been evaluated. 
The contributions of FCNC are harmless  
to the $K^0$-$\overline{K}^0$ and $D^0$-$\overline{D}^0$ mixings.
On the other hand, the elements related to the top-quark have 
sizable values 
$(C_R^u)_{tc}=0.000709$ and $(C_R^u)_{tu}=0.000284$.

The present model which can successfully give quark mass ratios 
and CKM matrix parameters  can also provide fruitful new physics. 
It seems that the model is worth investigating not only 
phenomenologically but also theoretically.
 
\vglue.2in

\centerline{\bf ACKNOWLEDGMENTS}

The author would like to thank H.~Fusaoka, T.~Kurimoto and M.~Tanimoto 
for helpful and enjoyable discussions. 
This work was supported by the Grant-in-Aid for Scientific Research, the 
Ministry of Education, Science and Culture, Japan (No.08640386). 


\vglue.4in

\centerline{\bf APPENDIX: STRUCTURE OF FCNC}
\vglue.05in

When the mass matrix $M$ given in (1.1) is transformed as 
$$
\overline{\psi_L} M \psi_R + h.c. = \overline{\psi'_L}D\overline{\psi'_R} + 
h.c. \ \ , 
\eqno(A1)
$$
where $\psi = (f, F)^T$, and $\psi' = U\psi$ is the mass-eigenstates, 
the vertex $\overline{\psi}_A\Gamma^{AB}\psi_B$  ($A, B= L, R$) is also 
transformed into $\overline{\psi}'_A\Gamma'^{AB}\psi'_B$, 
where 
$$
\Gamma'^{AB} = U_A \Gamma^{AB} U_B^\dagger \ \ . 
\eqno(A2)
$$
For simplicity, hereafter, we drop the indices $A, \ B$.
Correspondingly to (3.3), we denote the $6 \times 6$ matrix $\Gamma$ 
in terms of $3 \times 3$ matrices  $\Gamma_{ab}$  ($a, b=f,F)$ as 
$$
\Gamma = \left(\begin{array}{cc}
\Gamma_{ff} & \Gamma_{fF} \\ 
\Gamma_{Ff} & \Gamma_{FF} \\ 
\end{array} \right)  \ . 
\eqno(A3)
$$
Our interest is in the physical vertex $\Gamma'_{ff}$ which is 
given by 
$$
\Gamma'_{ff} = \sum_{a} \sum_{b} 
U_{fa}\Gamma_{ab} U_{fb}^\dagger \ \ , 
\eqno(A4)
$$
where $U_{ab}^\dagger \equiv (U_{ab})^\dagger = (U^\dagger)_{ba}$, 
because $(\Gamma'_{ff})_{ij}$ with $i \neq j$ mean transitions 
between $f_i$ and $f_j$, i.e.,  appearance of the FCNC. 

In our SU(2)$_L \times $SU(2)$_R \times $U(1)$_Y$ gauge model, 
the neutral currents  $J_L^\mu = g_L^Z \overline{\psi}\Gamma_L^\mu \psi$, 
which couple with the left-handed weak boson $Z_L^\mu$, 
are given by 
$$
\Gamma_L^\mu = \left(\begin{array}{cc}
c_L^f{\bf 1} & 0 \\ 
0 & c_L^F{\bf 1} \\ 
\end{array} \right) \cdot \frac{1}{2} \gamma^\mu (1-\gamma_5) + 
\left(\begin{array}{cc}
d_L^f{\bf 1} & 0 \\ 
0 & d_L^F{\bf 1} \\ 
\end{array} \right) \cdot \frac{1}{2} \gamma^\mu (1+ \gamma_5) \ \ , 
\eqno(A5)
$$
where 
$$
\begin{array}{rll}
c_L^f & = \pm \frac{1}{2} & - \sin^2\theta_L Q_f \ \ , \\ 
c_L^F & =  & -\sin^2\theta_L Q_F \ \ , \\ 
\end{array} 
\eqno(A6)
$$
$$
\begin{array}{rll}
d_L^f & = \pm\frac{1}{2}h_{L} & -\sin^2\theta_L Q_f \ \ , \\ 
d_L^F & =  & - \sin^2\theta_L Q_F \ \ , \\ 
\end{array} 
\eqno(A7)
$$
$$
\sin^2\theta_L = 1 - m_{W_L}^2/m_{Z_L}^2 \ \ , 
\eqno(A8)
$$
$$
h_{L} = -\frac{\sin^2\theta_L}{1 - \varepsilon/\cos^2\theta_L} 
\frac{\varepsilon}{\cos^2\theta_L} \ \ , 
\eqno(A9)
$$
$$
\varepsilon = m_{W_L}^2/m_{W_R}^2 \ \ , 
\eqno(A10)
$$
the factor $\pm\frac{1}{2}$ takes $+\frac{1}{2}$ and $-\frac{1}{2}$ for 
up- and down-fermions, respectively, and $Q_f \ (Q_F)$ is 
charge of the fermion $f$  $(F)$. 
Using the unitary condition for $U_{ab}$, $U_{ff}U_{ff}^\dagger + 
U_{fF}U_{fF}^\dagger = {\bf 1}$, we can express the physical vertex 
$\Gamma'_{Lff}$ as
$$
\begin{array}{c}
\Gamma^{'\mu}_{Lff} = \left(c_L^f U_{ff}^LU_{ff}^{L\dagger} + 
c_L^F U_{fF}^LU_{fF}^{L\dagger}\right) 
\cdot \frac{1}{2}\gamma^\mu(1 - \gamma_5) \\ 
+ \left(d_L^f U_{ff}^RU_{ff}^{R\dagger} + 
d_L^F U_{fF}^RU_{fF}^{R\dagger}\right) \cdot 
\frac{1}{2}\gamma^\mu(1 + \gamma_5) \\ 
= \left[c_L^f {\bf 1} - (c_L^f - c_L^F)U_{fF}^LU_{fF}^{L\dagger}
\right] \cdot 
\frac{1}{2}\gamma^\mu(1-\gamma_5) \\ 
+ \left[d_L^f {\bf 1} - (d_L^f - d_L^F)U_{fF}^RU_{fF}^{R\dagger}
\right] \cdot 
\frac{1}{2}\gamma^\mu (1 + \gamma_5) \ \ . \\ 
\end{array} \eqno(A11)
$$
Similarly, for the neutral current $J_R^\mu = g_R^Z \overline{\psi'} 
\Gamma_R^{'\mu} \psi'$, which couples with the right-handed weak boson 
$Z_L$, we obtain 
$$
\begin{array}{c}
\Gamma^{'\mu}_{Rff} = \left[c_R^f{\bf 1} - 
(c_R^f - c_R^F)U_{fF}^RU_{fF}^{R\dagger}\right] 
\cdot \frac{1}{2} \gamma^\mu(1 + \gamma_5) \\ 
+ \left[d_R^f {\bf 1} - (d_R^f - d_R^F) U_{fF}^LU_{fF}^{L\dagger}
\right] \cdot 
\frac{1}{2}\gamma^\mu(1-\gamma_5) \ \ , \\ 
\end{array} \eqno(A12)
$$
where 
$$
\begin{array}{rll}
c_R^f & = \pm \frac{1}{2} & - \sin^2\theta_R Q_f \ \ , \\ 
c_R^F & =  & - \sin^2\theta_RQ_F \ \ , \\ 
\end{array} \eqno(A13)
$$
$$
\begin{array}{rll}
d_R^f & = \pm \frac{1}{2}h_{R} & -\sin^2\theta_R Q_f \ \ , \\
d_R^F & =  & -\sin^2\theta_RQ_F \ \ , \\
\end{array} \eqno(A14)
$$
$$
\sin^2\theta_R = 1 - m_{W_R}^2/m_{Z_R}^2 \ \ , 
\eqno(A15)
$$
$$
h_{R} = -\frac{\sin^2\theta_R}{1 - \varepsilon\cos^2\theta_R} \ \ , 
\eqno(A16)
$$
$$
g_R^Z = -g_L^Z\frac{\sin\theta_L}{\sin\theta_R\cos\theta_R} 
\sqrt{\frac{1 - \varepsilon\cos^2\theta_R}
{1 - \varepsilon/\cos^2\theta_L}}
$$
$$
=\frac{e}{\cos\theta_L\sin\theta_R\cos\theta_R} 
\sqrt{\frac{1 - \varepsilon\cos^2\theta_R}
{1 - \varepsilon\cos^2\theta_R/\cos^2\theta_L}} \ \ . 
\eqno(A17)
$$

Thus, the FCNC are induced by the second terms 
$U_{fF}U_{fF}^\dagger$ 
with magnitude $(c^f - c^F)$ $[(d^f - d^F)]$. 
Therefore, we have denoted these matrices 
as $C_L^f\equiv U_{fF}^LU_{fF}^{L\dagger}$ 
and $C_R^f\equiv U_{fF}^RU_{fF}^{R\dagger}$ in (4.1).


\vglue.4in
\newcounter{0000}
\centerline{\large\bf References}
\begin{list}
{[~\arabic{0000}~]}{\usecounter{0000}
\labelwidth=0.8cm\labelsep=.1cm\setlength{\leftmargin=0.7cm}
{\rightmargin=.2cm}}
\item Y.~Koide and H.~Fusaoka, Z.~Phys. {\bf C71}, 459 (1966).
\item T.~Morozumi, T.~Satou, M.~N.~Rebelo and M.~Tanimoto, 
Preprint HUPD-9704 (1997), hep-ph/9703249.
\item Z.~G.~Berezhiani, Phys.~Lett.~{\bf 129B}, 99 (1983);
Phys.~Lett.~{\bf 150B}, 177 (1985);
D.~Chang and R.~N.~Mohapatra, Phys.~Rev.~Lett.~{\bf 58}, 1600 (1987); 
A.~Davidson and K.~C.~Wali, Phys.~Rev.~Lett.~{\bf 59}, 393 (1987);
S.~Rajpoot, Mod.~Phys.~Lett. {\bf A2}, 307 (1987); 
Phys.~Lett.~{\bf 191B}, 122 (1987); Phys.~Rev.~{\bf D36}, 1479 (1987);
K.~B.~Babu and R.~N.~Mohapatra, Phys.~Rev.~Lett.~{\bf 62}, 1079 (1989); 
Phys.~Rev. {\bf D41}, 1286 (1990); 
S.~Ranfone, Phys.~Rev.~{\bf D42}, 3819 (1990); 
A.~Davidson, S.~Ranfone and K.~C.~Wali, 
Phys.~Rev.~{\bf D41}, 208 (1990); 
I.~Sogami and T.~Shinohara, Prog.~Theor.~Phys.~{\bf 66}, 1031 (1991);
Phys.~Rev. {\bf D47}, 2905 (1993); 
Z.~G.~Berezhiani and R.~Rattazzi, Phys.~Lett.~{\bf B279}, 124 (1992);
P.~Cho, Phys.~Rev. {\bf D48}, 5331 (1994); 
A.~Davidson, L.~Michel, M.~L,~Sage and  K.~C.~Wali, 
Phys.~Rev.~{\bf D49}, 1378 (1994); 
W.~A.~Ponce, A.~Zepeda and R.~G.~Lozano, 
Phys.~Rev.~{\bf D49}, 4954 (1994).
\item M.~Gell-Mann, P.~Rammond and R.~Slansky, in {\it Supergravity}, 
edited by P.~van Nieuwenhuizen and D.~Z.~Freedman (North-Holland, 
1979); 
T.~Yanagida, in {\it Proc.~Workshop of the Unified Theory and 
Baryon Number in the Universe}, edited by A.~Sawada and A.~Sugamoto 
(KEK, 1979); 
R.~Mohapatra and G.~Senjanovic, Phys.~Rev.~Lett.~{\bf 44}, 912 (1980).
\item CDF Collaboration, F.~Abe {\it et al.}, Phys.~Rev.~Lett. 
{\bf 73}, 225 (1994).
\item N.~Cabibbo, Phys.~Rev.~Lett.~{\bf 10}, 531 (1996); 
M.~Kobayashi and T.~Maskawa, Prog.~Theor.~Phys.~{\bf 49}, 652 (1973).
\item J.~C.~Pati and A.~Salam, Phys.~Rev. {\bf D10}, 275 (1974); 
R.~N.~Mohapatra and J.~C.~Pati, Phys.~Rev. {\bf D11},
 366 and 2588 (1975);
G.~Senjanovic and R.~N.~Mohapatra, Phys.~Rev. {\bf D12}, 1502 (1975).
\item T.~Kurimoto, A.~Tomita and S.~Wakaizumi, Phys.~Lett. 
{\bf B381}, 470 (1996).
\item Y.~Koide, Preprint US-96-09 (1996), hep-ph/9701261 
(unpublished).
\item H.~Terazawa, University of Tokyo, Report No.~INS-Rep.-298 
(1977) (unpublished): 
H.~Harari, H.~Haut and J.~Weyers, 
Phys.~Lett.~{\bf 78B}, 459 (1978);
T.~Goldman, in {\it Gauge Theories, Massive Neutrinos and 
Proton Decays}, edited by A.~Perlumutter (Plenum Press, New York, 
1981), p.111;
T.~Goldman and G.~J.~Stephenson,~Jr., Phys.~Rev.~{\bf D24}, 236 (1981); 
Y.~Koide, Phys.~Rev.~Lett. {\bf 47}, 1241 (1981); 
Phys.~Rev.~{\bf D28}, 252 (1983); {\bf 39}, 1391 (1989);
C.~Jarlskog, in {\it Proceedings of the International Symposium on 
Production and Decays of Heavy Hadrons}, Heidelberg, Germany, 1986
edited by K.~R.~Schubert and R. Waldi (DESY, Hamburg), 1986, p.331;
P.~Kaus, S.~Meshkov, Mod.~Phys.~Lett.~{\bf A3}, 1251 (1988); 
Phys.~Rev.~{\bf D42}, 1863 (1990);
L.~Lavoura, Phys.~Lett.~{\bf B228}, 245 (1989); 
M.~Tanimoto, Phys.~Rev.~{\bf D41}, 1586 (1990);
H.~Fritzsch and J.~Plankl, Phys.~Lett.~{\bf B237}, 451 (1990); 
Y.~Nambu, in {\it Proceedings of the International Workshop on 
Electroweak Symmetry Breaking}, Hiroshima, Japan, (World 
Scientific, Singapore, 1992), p.1.
\item Y.~Koide,  Mod.~Phys.~Lett.{\bf A8}, 2071 (1993).
\item Y.~Koide and H.~Fusaoka, Prog.~Theor.~Phys. {\bf 97}, 459 (1966).

\end{list}

\end{document}